# Long Time Response of Aging Glassy Polymers

Yogesh M Joshi,

Department of Chemical Engineering, Indian institute of Technology Kanpur,

Kanpur 208016, India

Email: joshi@iitk.ac.in

**Abstract:**

Aging amorphous polymeric materials undergo free volume relaxation, which causes slowing down of the relaxation dynamics as a function of time. The resulting time dependency poses difficulties in predicting their long time physical behavior. In this work, we apply effective time domain approach to the experimental data on aging amorphous polymers and demonstrate that it enables prediction of long time behavior over the extraordinary time scales. We demonstrate that, unlike the conventional methods, the proposed effective time domain approach can account for physical aging that occurs over the duration of the experiments. Furthermore, this procedure successfully describes time – temperature superposition and time – stress superposition. It can also allow incorporation of varying dependences of relaxation time on aging time as well as complicated but known deformation history in the same experiments. This work strongly suggests that the effective time domain approach can act as an important tool to analyze the long time physical behavior of aging amorphous polymeric materials.

# I Introduction:

Glassy materials such as amorphous polymers (Hodge 1995), colloidal glasses and gels (Fielding *et al.* 2000), spin glasses (Rodriguez *et al.* 2003) are usually time dependent materials owing to thermodynamically out of equilibrium state they are arrested in. Typically specific volume (and specific entropy/enthalpy) associated with an amorphous polymer is in excess compared to its equilibrium value (Donth 2001, Larson 1999, Struik 1978). Inherent tendency of any material to achieve equilibrium initiates structural rearrangement in the amorphous polymeric compounds so as to cause relaxation of specific volume (accompanied by decrease in specific entropy/enthalpy) as a function of time (Callen 1985, Struik 1978). This phenomenon is addressed in the literature as physical aging (Hodge 1995). In polymeric glasses, while taking the material closer to the equilibrium state, physical aging manifests itself by causing enhancement in relaxation time and stiffness as a function of time (Struik 1978). This time dependency continuously changes the material response to the external stimuli such as application of deformation field (stress field or strain field), electric field, etc. Consequently, a priory prediction of the long time behavior of the response function becomes a challenging task. In this work we employ the so called effective time domain approach, which allows direct prediction of long time material response of amorphous polymers by carrying out tests over the practical timescales.

The amorphous polymeric materials when quenched rapidly from the molten state undergo glass transition (McKenna 1994). Physical aging initiates in the material subsequent thermal quench (Hodge 1995). The time elapsed since thermal quench is typically known as the aging time or the waiting time. If the material is deformed during this annealing period, response functions (such as creep compliance and relaxation modulus, depending upon the nature of the deformation field) not only depend on the time passed since imposition of deformation field but also on the aging time (Fielding *et al.* 2000). In addition, the response function is strongly affected by aging that occurs over the duration of imposition of deformation field. Both these effects contribute to the degree of difficulty in making the prediction of long time behavior.

Historically Struik was the first to address this problem, who carried out enormous amount of experimental and theoretical work on this subject (Struik 1978). He proposed an effective time methodology to predict long time behavior of response function from a 'momentary' time – aging time superposition obtained by carrying out short time tests. The rationale behind employing only the short time experimental data (data obtained over time much smaller than the aging time at which experiments were started) was to avoid non-linarites originating from the aging that occurs during the course of experiments. The framework of 'momentary' time approach was adopted by subsequent studies that attempted to predict long time behavior of variety of different amorphous polymeric materials, semicrystalline polymers, polymer composites, polymers with different thermal histories, etc. (Arnold and White 1995, Bradshaw and Brinson 1997, Guo and Bradshaw 2009, Guo *et al.* 2009, Nicholson *et al.* 2000, Read *et al.* 1990, Tomlins *et al.* 1994, Zheng and Weng 2002). Effect of temperature and deformation field on relaxation time and its variation as a function of time led to the time – aging time – temperature and time – aging time – stress/strain superpositions (Dorigato *et al.* 2010, Guo and Bradshaw 2009, Jazouli *et al.* 2005, Kolařik 2003, Kolarik and Pegoretti 2006, O'Connell and McKenna 1997, O'Connell and McKenna 2002, Schoeberle *et al.* 2008, Shi *et al.* 2005, Struik 1978). Struik also proposed an empirical dependence of the momentary creep compliance on time using a stretched exponential function. If such a function is fitted to the available momentary experimental data, the predictive capacity of the momentary approach has no limit, subject to the validity of assumption that the experimental data does strictly follow this relation. Remarkably, the application of the Struik methodology was not just limited to polymeric glasses, but it was successfully demonstrated for soft glassy materials and spin glasses as well (Cloitre *et al.* 2000, McKenna *et al.* 2009, Negi and Osuji 2009, Reddy and Joshi 2008, Sibani and Kenning 2010). Interestingly time– temperature and time – stress superposition were also found to be valid for soft glassy materials (Awasthi and Joshi 2009, Joshi and Reddy 2008, Shaukat *et al.* 2010). These momentary superpositions aid the prediction of long time behavior even further.

Owing to its significant potential, the Struik methodology is universally accepted in the polymer community. However, it suffers from some important limitations that impede its predictive capacity. Firstly this methodology allows consideration of only that duration data, which is much shorter than the aging time (typically 10 % of the aging time). Therefore, this method cannot take advantage of the availability or measurability of the experimental data over longer durations. In addition, even consideration of data over duration typically 10 % (or lesser) of the aging time cannot ensure the absence of aging over the duration of an experiment thereby inducing errors in the prediction. In order to address these limitations Shahin and Joshi (Shahin and Joshi 2011) proposed an effective time domain approach based on the theoretical developments by Hopkins (Hopkins 1958), Struik (Struik 1978) and Fielding and coworkers (Fielding *et al.* 2000) and successfully applied the same to various soft glassy materials. This approach allows time – aging time superposition to consider any duration of process data (Shahin and Joshi 2012). Gupta and coworkers (Gupta *et al.* 2012) included the effect of temperature, and for the first time proposed a systematic treatment of time – temperature superposition for aging time dependent soft materials. This facilitated prediction of rheological behavior for glassy soft materials over longer times. Baldewa and Joshi (Baldewa and Joshi 2012), on the other hand, extended this approach to include the effect of stress on the aging behavior. In this work, we apply this effective time domain approach to experiments on the aging amorphous polymeric systems reported by Struik (Struik 1978) and by Read and Tomlins (Read and Tomlins 1997). We also apply time – temperature superposition and time stress superposition in the effective time domain to Struik's data. We demonstrate that the effective time domain approach not just allows predictions over the exceptionally large durations but it also shows improvement in the same. In the next section, we briefly discuss the effective time domain approach and Struik's momentary experiments approach. Subsequently, we discuss application of effective domain approach to the experimental data.

**II. Effective time theory**

The concept of effective time is originally due to Hopkins (Hopkins 1958). Below we present two approaches namely the effective time domain approach and Struik's 'momentary' experiments approach, available in the literature to predict the long time creep behavior. Although the latter approach is a special case of the former approach, we present the original methodology suggested by Struik (Struik 1978). We represent the discussion mentioned below in terms of creep experiments, wherein stress was applied at time $t_w$ elapsed since thermal quench (aging time), while $t$ is the total time. The equations mentioned below can be equivalently expressed to represent stress relaxation following step strain as described elsewhere (Gupta *et al.* 2012).

A. <u>Effective time domain approach</u>

The equilibrium soft materials are known to follow the time translational invariance (TTI), which implies that in a constitutive equation replacement of time $t$ by $t+c$, where $c$ is a constant, leaves the nature of the constitutive equation unchanged (Denman 1968). Consequently the materials that follow TTI also follow the Boltzmann superposition principle (BSP). The most conventional form of BSP, wherein the stress is an independently controlled variable, is given by (Bird *et al.* 1987):

$$\gamma(t) = \int_{-\infty}^{t} J(t - t_w) \frac{d\sigma}{dt_w} dt_w, \qquad (1)$$

where $\gamma$ is strain induced in the material at time $t$ in response to stress $\sigma$ imposed on the material at time $t_w$ and $J$ is creep compliance. It should be noted that, owing to the applicability of TTI, $J$ is only a function of creep time (time passed since the imposition of the stress field): $J = J(t - t_w)$. Consequently the creep data collapses to form a master curve when plotted as a function of creep time $(t - t_w)$ irrespective of $t_w$ at which the creep stress was applied.

For aging materials with inherent time dependency, TTI is not applicable as properties of the material are different at $t + c$ from that at $t$. Consequently the conventional form of the Boltzmann superposition principle, wherein compliance depends only on $t - t_w$ or creep time, is not applicable. Under such conditions, compliance shows an additional dependence on time at which deformation field is applied ($t_w$) (Fielding *et al.* 2000). Accordingly, the general expression of Boltzmann superposition principle can be written as:

$$\gamma(t) = \int_{-\infty}^{t} J(t - t_w, t_w) \frac{d\sigma}{dt_w} dt_w. \qquad (2)$$

This additional dependence of compliance on aging time ($t_w$) is the main obstacle in extending the linear viscoelastic analysis to aging glassy materials. In glassy materials the characteristic relaxation time depends on the time elapsed since thermal quench (aging time, $t_w$): $\tau = \tau(t_w)$. Therefore, in order to eliminate this additional dependence of $J$ on $t_w$, it is suggested that Boltzmann superposition principle be represented in the effective time domain rather than in the real time domain (Shahin and Joshi 2011, Struik 1978). The effective time is obtained by normalizing the real time by the time dependent relaxation time such that time dependency ceases in the effective time domain. The effective time is defined as (Shahin and Joshi 2011):

$$\xi(t) = \int_0^t \tau_0 dt'/\tau(t'),  \qquad (3)$$

where $\tau_0$ is a constant relaxation time associated with the effective time scale $\xi$. In the $\xi$ - domain, the relaxation time remains constant, consequently $J$ in BSP is only a function of the effective time passed since the application of stress field: $J = J\big(\xi(t) - \xi(t_w)\big)$. The modified Boltzmann superposition principle is given by (Shahin and Joshi 2011):

$$\gamma(t) = \int_{-\infty}^t J\big(\xi(t) - \xi(t_w)\big) \frac{d\sigma}{dt_w} dt_w \qquad (4)$$

This suggests that if compliance is plotted against $\xi(t) - \xi(t_w)$, the data is expected to collapse onto a master curve irrespective of the time at which creep experiments were started and the duration of the same. When relaxation time is independent of real time ($\tau(t)$=constant), equation (4) reduces to the conventional form of Boltzmann superposition principle for the equilibrium soft materials given by equation (1).

For the glassy materials in general and the polymer glasses in particular relaxation time is observed to show power law dependence of on real time given by (Cloitre *et al.* 2000, Derec *et al.* 2003, Fielding *et al.* 2000, McKenna *et al.* 2009, O'Connell and McKenna 1997, Struik 1978, Tomlins and Read 1998):

$$\tau(t) = A\tau_m^{1-\mu}t^{\mu}, \tag{5}$$

where $A$ is a constant, $\tau_m$ is microscopic relaxation time and $\mu$ $\left(= d\ln\tau/d\ln t\right)$ is shift rate (Struik 1978). The effective time passed since imposition of stress field can then be represented by incorporating equation (5) into equation (3) to give:

$$\xi(t) - \xi(t_w) = \int_{t_w}^{t}\tau_0 dt'/\tau(t') = \frac{\tau_0\tau_m^{\mu-1}}{A}\left[\frac{t^{1-\mu} - t_w^{1-\mu}}{1-\mu}\right] \tag{6}$$

Joshi and coworkers employed effective time domain approach to predict the long time creep and stress relaxation behavior in variety of soft non-ergodic materials (Shahin and Joshi 2011, Shahin and Joshi 2012). In conventional practice consideration of effect of temperature on glassy materials while obtaining a superposition is complicated as relaxation time depends on temperature as well as aging time. However, in the effective time domain since relaxation time is constant for a given temperature, application of time – temperature superposition becomes as straightforward as application of the same to the equilibrium materials in the real time domain. Therefore, effective time domain approach provides the *only* systematic way to apply time – temperature superposition to the time dependent materials. In addition to aging time and temperature, magnitude of creep stress also affects the relaxation time. Incorporation of the same leads to the time – aging time – stress superposition in the $\xi$ - domain (Baldewa and Joshi 2012, Gupta *et al.* 2012). Both these techniques were observed to facilitate the prediction of long time creep behavior in the glassy materials.

B.  Struik's 'momentary' experiments approach (Struik 1978):

For amorphous polymers in glassy state, physical aging is accompanied by free volume relaxation leading to progressive slowing down of the segmental mobility as a function of time lapsed since thermal quench ($t_w$). Struik suggested that during an interval $t_w$ and $t'_w$ ($> t_w$), segmental mobility $M(t_w)$ decreases by a factor $a$ given by: $a(t'_w, t_w) = M(t'_w)/M(t_w)$. Struik considered the mobility to have a power law dependence on the aging time, leading to: $a(t'_w, t_w) = (t_w/t'_w)^\mu$. He proposed that, in a creep experiment at time $t$ (time $t - t_w$ after $t_w$ at which stress field was applied), time required to complete a certain relaxation process for a material over a duration of $dt$, is related to $d\lambda$, if material retained the same relaxation time that at $t = t_w$. The relation between $d\lambda$ and $dt$ is given by: $d\lambda = a(t,t_w)dt$. He represented timescale $\lambda$ as an effective time scale defined as (Struik 1978):

$$\lambda(t,t_w) = \int_{t_w}^{t} a(t',t_w)dt' \tag{7}$$

It is assumed that for short duration or 'momentary' experiments, wherein creep time is much smaller than aging time ($t - t_w << t_w$), aging that occurs over the experimental time scales can be neglected. Consequently, relaxation behavior over the experimental timescales can be considered to be the same as that of at the beginning of creep ($t_w$). Struik carried out creep experiments on a variety of aging polymeric materials at different aging times ($t_w$), different temperatures and different stress fields over durations $(t - t_w)$ much smaller than $t_w$. The representative nature of creep curves obtained at different $t_w$ and temperatures are schematically illustrated in figure 1. Struik observed that the experimentally obtained creep curves have self-similar curvatures in the limit $t - t_w << t_w$. Consequently by appropriately shifting the creep curves in the 'momentary' domain a master curve was obtained as shown in figure 1. Similar to temperature, momentary creep curves at different stresses but same temperature and aging time also can be shifted to obtain a momentary superposition. The qualitative nature of such shifting procedure is same as that described in figure 1. Struik proposed that for an experiment started at aging time $t_w$, compliance induced at time $t$ ($t - t_w$ after time $t$) is same as that in the momentary experiments at effective time $\lambda(t, t_w)$ so that compliance is given by (Struik 1978),

$$J_{t_{w1}}(t - t_{w1}) = J_{t_{w1}}(\lambda). \tag{8}$$

Equation (8) can be used to transform the momentary master curve to predict the creep behavior in the real time domain, where $\lambda$ is given by equation (7), as described schematically in figure 1.

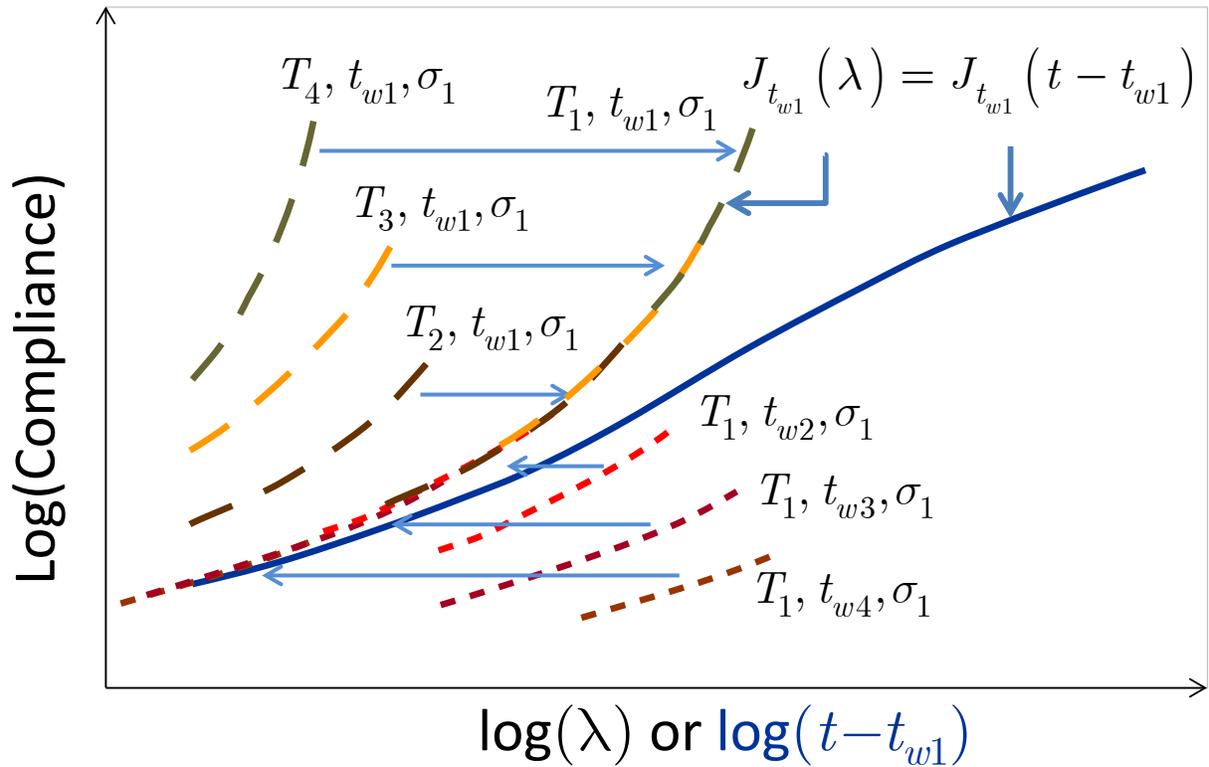

**Figure 1.** Schematic description of Struik protocol. The momentary creep curves ($t - t_w \ll t_w$) at different waiting times (dotted lines) and temperatures (dashed lines) demonstrate self-similar curvature. The horizontal and vertical shifting of these curves lead to momentary superposition as shown. The momentary superposition can be transformed to get long time creep prediction shown by dark blue line using equations (7) and (8). To avoid crowding we have not shown schematic creep curves at higher stresses but same temperature and aging time. The qualitative nature of creep curves with different stresses is observed to be same as that of at different temperatures shown here.

The concept of effective time is also present in Tool - Narayanswamy – Moynihan (TNM) (Narayanswamy 1971) and Kovacs–Aklonis–Hutchinson–Ramos (KAHR) (Kovacs et al. 1979) approaches, which relate change in specific volume in a glassy material subjected to time dependent change in temperature (Hodge 1994, Struik 1978). The resulting expression of these approaches is equivalent to the Boltzmann superposition principle [strain and stress in equation (1) are replaced by the specific volume and temperature respectively], when written in an effective time domain. In this case effective time is obtained by normalizing real time by relaxation time that depends on time through its dependence on specific volume and temperature. However, the approaches by TNM and KHAR do not account for the effect of deformation field.

Many variants of the Struik's procedure have been reported in the literature that are specific to particular material behavior or to the applied thermal/stress field. Zheng and Weng (Zheng and Weng 2002) modified the Struik's theory to analyze chrono-rheologically simple materials whose long term creep response shows self-similar curvature to yield a superposition by horizontal shifting. Tomlins and coworkers (Read et al. 1992, Read and Tomlins 1997, Tomlins and Read 1998, Tomlins et al. 1994) analyzed long time behavior of aging polymeric materials by fitting a stretched exponential function with embedded effective time to account for aging during the course of creep. Their model also considered stress down jumps in a similar fashion as stress up jumps, which led to fine prediction of not just two-step creep but also that of strain recovery after the stress was removed. In a recent work, Guo and Bradshaw (Guo and Bradshaw 2009) subjected the aging sample to non-isothermal history before performing the long term creep experiments under isothermal conditions. In this case owing to non-isothermal history, the shift rate $\mu$ of the material was not constant but depended on time $\mu^*(t')$, which they obtained from the modified KHAR model predictions. They estimated the effective time $\lambda$ from equation (7) but with $\mu$ replaced by $\mu^*$. They divided the measured creep data into two time zones: the short and the long time part of the data. They fitted the stretched exponential $D(t) = D_0 \exp(\lambda/\tau)^\beta$ to the short time creep data to obtain the fitting parameters and therefore the momentary curve as shown in figure 1. Finally, by using equation (8) they predicted the long term data which matched with the experimentally obtained data very well. The methodology presented in the present work, however, differs from the approach of Guo and Bradshaw since it can accommodate aging that occurs during the course of creep in the effective time domain. Consequently it does not treat the short time and long time creep data separately, which is a fundamental feature of the Struik protocol that is employed by Guo and Bradshaw. Secondly our approach does not use any kind of fitting relation such as stretched exponential as it may induce errors as discussed in the introduction section.

**IV. Analysis of the experimental data**

In this section we apply the effective domain approach to experimental long time aging data available in the literature. Figure 2 describes the data (obtained from figure 108 of Struik (Struik 1978)) for long time torsional creep behavior of rigid Poly Vinyl Chloride (PVC) measured at different aging times after quenching the material from 90°C to various temperatures mentioned in figure 2. It can be seen that compliance induced in the material is less for experiments started at higher aging time for a given temperature. On the other hand, compliance shows substantial increase with temperature. For all the data reported in figure 2, the time over which the creep experiments were carried out is much greater than the aging time at which experiments were started ($t - t_w >> t_w$). Under such conditions, significant part of the aging occurs during the creep experiment itself. Consequently, as evident from figure 2, the curvatures of the creep curves are not self-similar when plotted as a function of real time. Therefore, compliance cannot be superposed by mere horizontal shifting. In order to apply the Struik's procedure, horizontal shifting of only the limited creep data can be carried out (in the limit: $t - t_w << t_w$), thereby not utilizing the information available over the remaining duration of the experiment. According to the effective time domain approach, we can transform the entire data from the real time domain to the effective time domain, wherein compliance is only a function of the effective time lapsed since the imposition of stress field $[J = J(\xi(t) - \xi(t_w))]$. In figure 3 we plot compliance as a function of $[t^{1-\mu} - t_w^{1-\mu}]/(1-\mu)$ (which is nothing but $\left[\xi(t) - \xi(t_w)\right] A / (\tau_0 \tau_m^{\mu-1})$ according to equation (6)). It can be seen that creep compliance shows remarkable superpositions in the effective time domain. This validates the applicability of the Boltzmann superposition principle. The values of $\mu$, for which superpositions are obtained increase with the increase in temperature over the explored range of temperatures as shown in the inset of figure 3.

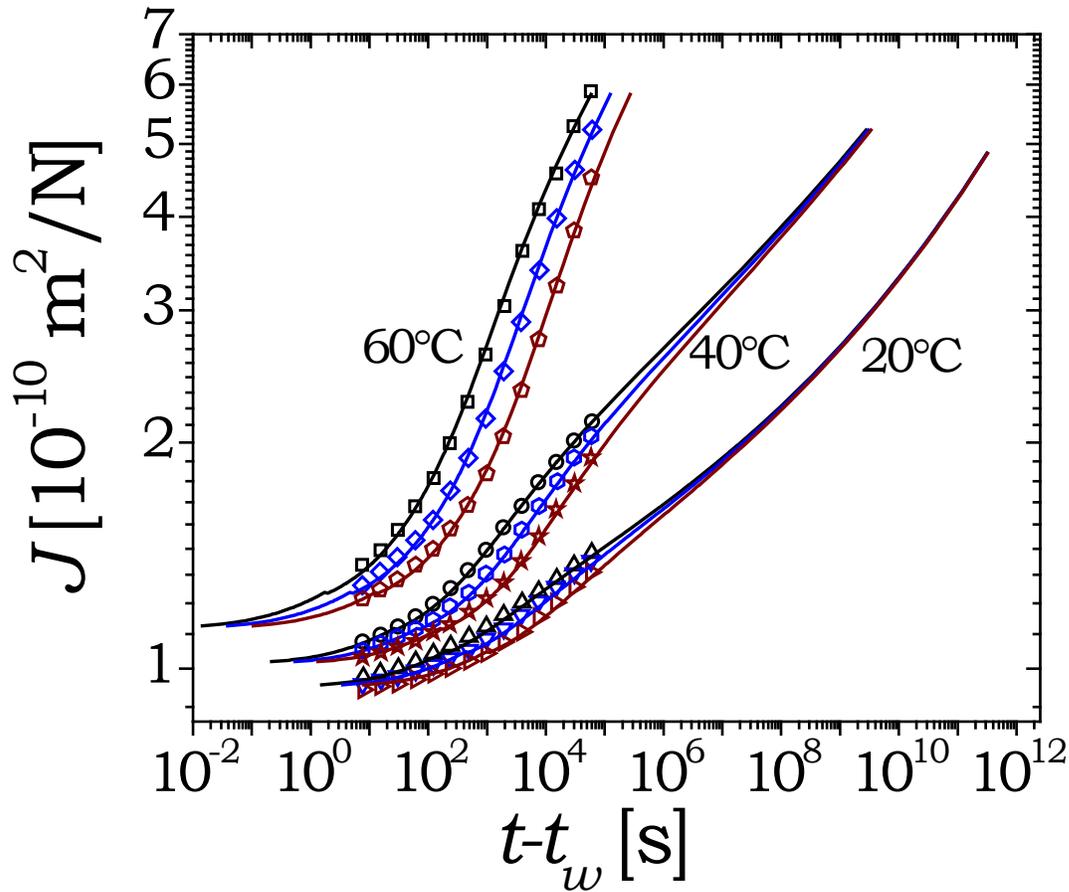

**Figure 2.** Long time torsional creep data of rigid PVC obtained from Struik (Figure 108) (Struik 1978) at various temperatures and aging times (in respective groups from top to bottom: 0.33 h (black), 1 h (blue), 3 h (wine)). Lines are the effective time domain theory predictions of long and short time behavior obtained by applying equation (8) to the superposition shown in figure 4.

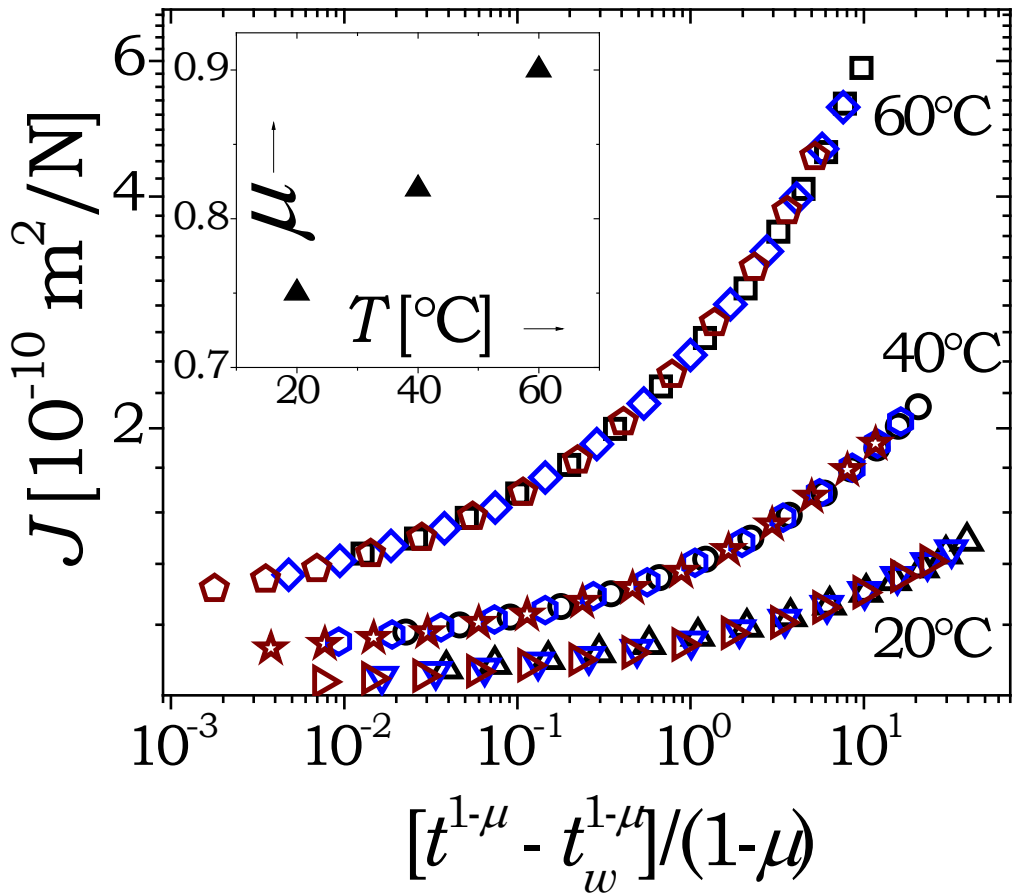

**Figure 3.** Time – aging time superposition of the creep data plotted in figure 2 represented in the effective time domain at respective temperatures. The inset shows corresponding values of shift rate $\mu$ as a function of temperature needed to obtain the superposition.

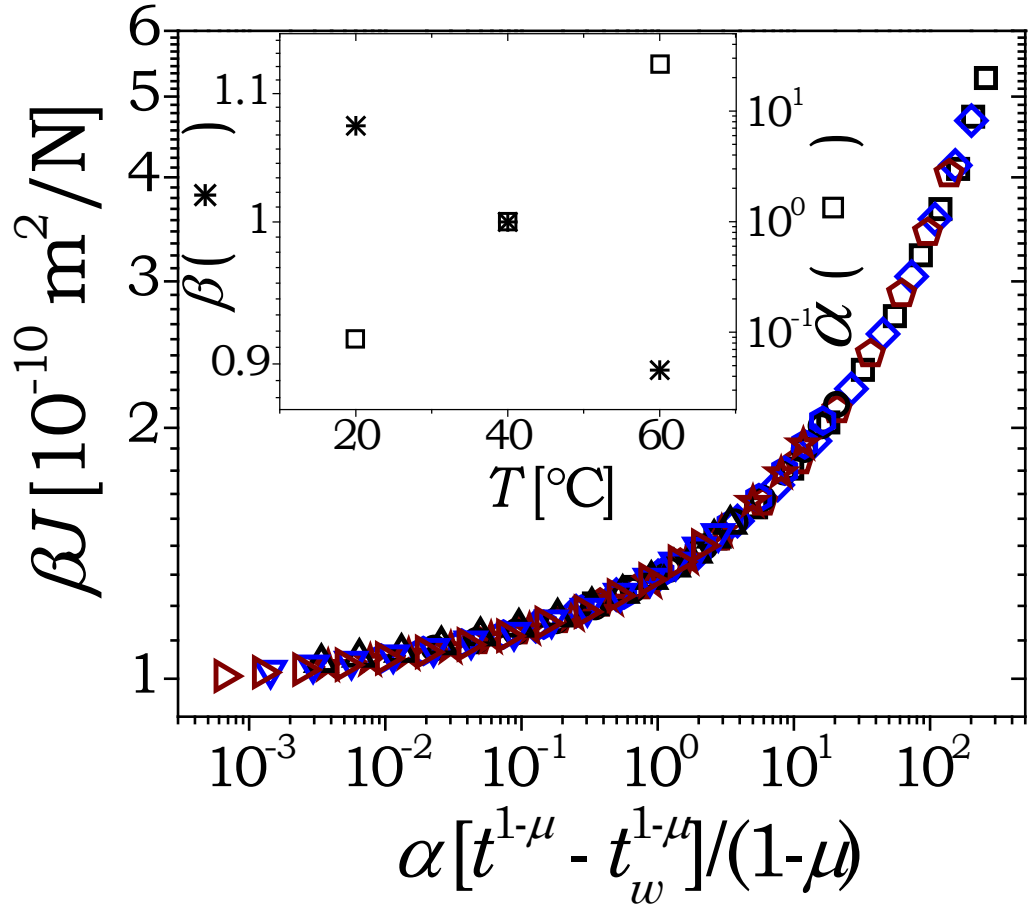

**Figure 4.** Time – aging time – temperature superposition obtained by horizontally and vertically shifting the individual time – aging time superpositions at respective temperatures shown in figure 3. Variation of vertical and horizontal shift factors is shown in the inset.

The individual superpositions at various temperatures shown in figure 3 can be seen to have self-similar curvature. In the effective time domain, each of these superpositions is characterized by a constant relaxation time which can be considered to depend only on temperature. This scenario is then equivalent to having rheological data of an equilibrium soft material at different temperatures in the real time domain. Therefore horizontal and vertical shifting of these superpositions in the effective time domain, shown in figure 3, leads to a comprehensive superposition, which we plot in figure 4. Existence of such superposition suggests that change in temperature as well as physical aging affects only the average value of relaxation time and not the shape of relaxation time spectrum (Struik 1978). It also suggests that $\alpha$ – time scale and $\beta$ – timescale are well separated, as usually is the case with molecular glasses (McKenna *et al.* 2009). The vertical and horizontal shift factors describe the dependence of the modulus and relaxation time on temperature respectively. It should be noted that comprehensive superposition shown in figure 4 is plotted in an effective time domain, where we consider relaxation time to have a constant value $\tau_0$ associated with the reference temperature. Therefore, in order to have the abscissa of figure 4 to be represented by $\left[\xi(t) - \xi(t_w)\right]/\tau_0$, the horizontal shift factor must be $\alpha = \tau_m^{\mu-1}/\tau_{mR}^{\mu_R-1}$, where subscript $R$ represents the reference state. Figure 4 suggests that evolution of compliance as a function of normalized effective time at the reference temperature ($T_R$=40°C) will follow the path traced by comprehensive superposition. Therefore inverting superposition from the effective time domain to the real time domain will facilitate the prediction of very long as well as very short time creep data. Similar comprehensive superposition can also be obtained at the other two temperatures as well. Given that at the reference temperature $\alpha = 1$, the abscissa of figure 4 given by: $\theta = \left(t^{1-\mu} - t_w^{1-\mu}\right)/(1-\mu)$ can be inverted from the effective time domain to the real time domain as (Shahin and Joshi 2011),

$$t - t_w = \left\{ \theta(1-\mu) + t_w^{1-\mu} \right\}^{1/(1-\mu)} - t_w. \tag{9}$$

We use equation (8) to transform the comprehensive superposition from the effective time domain to the real time domain for all the temperatures and the aging times. We plot the resulting creep curves as lines in figure 2. It can be seen that the creep curves at higher temperatures and smaller aging times enable prediction of the long time creep behavior at lower temperatures and higher aging times. On the other hand the creep behavior at lower temperatures and higher aging times aids prediction of the very small time creep behavior at the higher temperatures and smaller aging times. Figure 2 shows that the effective time procedure employed in this work allows a prediction of creep behavior at 20°C over six orders of magnitudes greater than the creep time. The prediction of small time behavior of creep data at 60°C is also important as instrument inertia usually does not allow exploration of small time-scales accurately. This example clearly shows a successful validation of time – temperature superposition in the effective time domain and its application in predicting the very long as well as the very short time creep behavior.

We believe that the time-scales over which the prediction of the low temperature data is demonstrated in figure 2, is far greater than what can be made with the Struik's momentary superposition procedure. This is due to the fact that momentary superposition procedure does not allow consideration of experimental data over those process times which are affected by aging. This significantly impedes the predictive capacity. As demonstrated in figures 2 to 4, we employ creep experiments started at the aging times between 1/3 h to 3 h over the process time of over 27 hours to predict the long time behavior. The present procedure facilitates the prediction over the extraordinary time scales by using $\xi$ time domain approach. In addition, it should be noted that this procedure does not assume any empirical dependence of creep compliance on time, as is the case in Struik's momentary experiments approach.

Next, we apply this procedure to short time creep data available at different temperatures as shown in figure 5(a). This data, which is also due to Struik (Struik 1978) (Figure 103), is generated at fixed aging time ($t_w$ = 2 h), by quenching rigid PVC from its melt state to respective temperatures over the range: 70 to 20°C. Time dependent evolution of relaxation time at different temperatures is characterized by different values of $\mu$. In figure 5(b) we plot creep curves in the effective time domain which demonstrate superposition after appropriate vertical and horizontal shifting. In order to plot the data in the effective time domain, we have used values of $\mu$ in the range 0.8 to 0.9, which are very close to that suggested by Struik. The horizontal and vertical shift factors employed in figure 5(b) are plotted as a function of temperature as an inset in the same figure. For the same system, Struik also obtained the long time creep data at 40°C for two aging times 0.5 h and 15.4 h, which is shown in figure 6. The black lines shown in figure 6 are the predictions of long time behavior obtained by inverting the superposition shown in figure 5(b) from the effective time domain to the real time domain using equation (8). It can be seen that the procedure proposed in the present work predicts the experimental data very well. In the same figure, we also plot the prediction of long time behavior obtained by Struik as dashed blue lines. Struik's prediction is based on horizontal and vertical shifting of the creep data in the real time domain and obtaining the long time behavior using equation (7). It can be seen that the effective time domain approach does a significantly better job in predicting the long time behavior than Struik's momentary experiments approach.

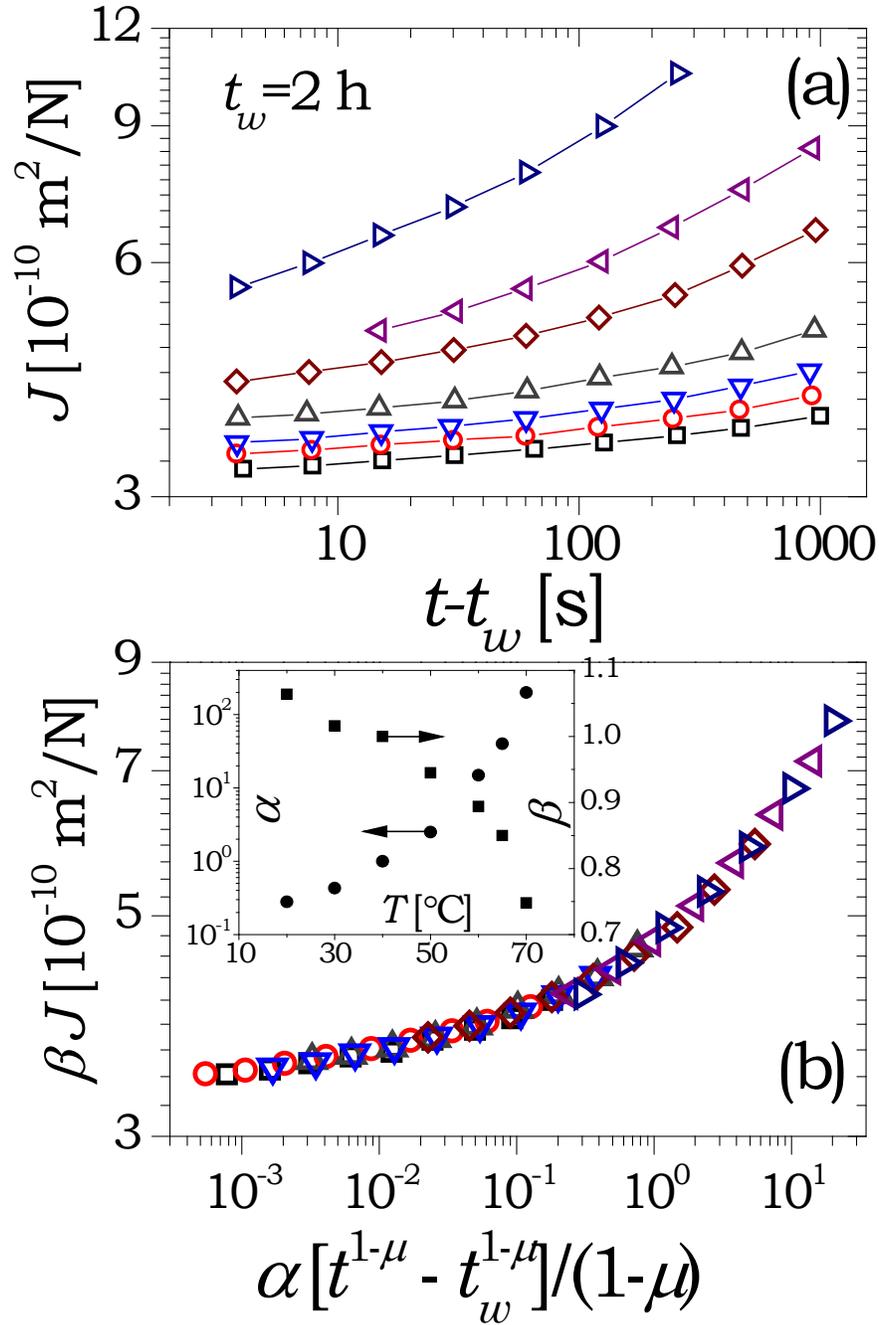

**Figure 5.** (a) Short time creep curves ($t_w$=2 h) for rigid PVC under tensile loading taken from Struik (Figure 103) (Struik 1978), from top to bottom: 70, 65, 60, 50, 40, 30, 20°C. (b) Representation of the data shown in the top figure in the effective time domain. Vertical and horizontal shift factors are shown in the inset. Values of $\mu$ at respective temperatures are very close to that suggested by Struik (Figure 112).

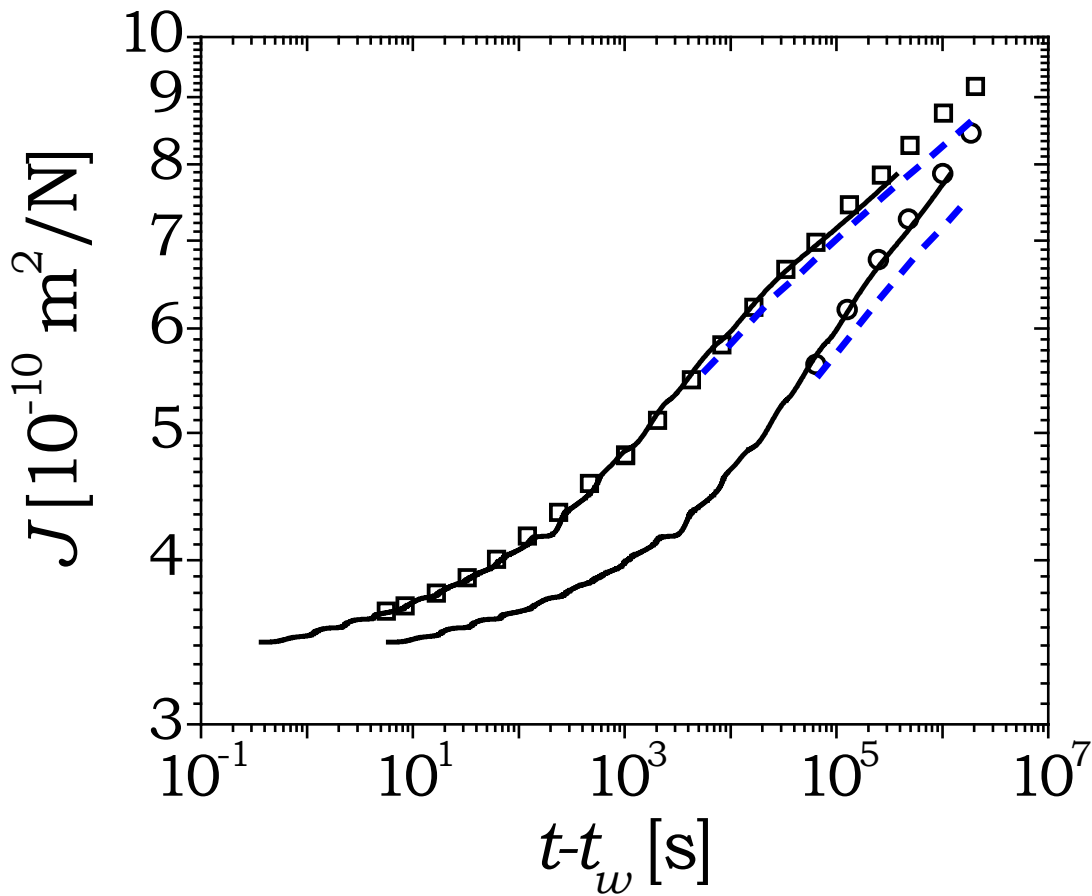

**Figure 6** The long time creep data of rigid PVC under tensile load at 40°C for (squares: $t_w$=0.5 h, circles: $t_w$=15.4 h) taken from Struik (figure 111) (Struik 1978). Black lines

show respective predictions of long time creep behavior from the effective time domain approach obtained by applying equation (8) to the superposition shown in figure 5(b). Dashed blue lines are the long time predictions obtained by Struik's procedure (Figure 111 (Struik 1978)).

We feel that the poor prediction with Struik's approach shown in figure 6 is due to an assumption that aging is absent over the duration of short time experiments. Different glassy systems are known to undergo aging at varying rates so that the absence of aging over the process time cannot be justified even though it is significantly smaller than the aging time, and the resultant superposition on real time scale appears to be fine. Consequently, transformation of such superposition to obtain data over the long duration induces errors. Effective time domain approach described in the present work, on the other hand, takes into account the effect of aging over the experimental time scales and therefore leads to better prediction. This example clearly suggests that even for analyzing short time creep data, use of effective time domain approach proves advantageous than the conventional approach.

The reason behind significant predictive capacity through time – aging time – temperature superposition is due to the strong influence of temperature on the relaxation time of the material. Since relaxation time decreases with temperature, processes that take very long time at low temperatures can be completed in very short time at a higher temperature. In addition to temperature, stress field is also known to decrease the relaxation time of the material. (Although effect of stress on the aging polymeric material is not as well-defined as that of temperature, we assume that the stress beyond a certain value indeed causes slowing down of the aging dynamics. For further discussion on this topic, refer to McKenna (McKenna 2003)) Tomlins and Read (Read and Tomlins 1997, Tomlins and Read 1998) carried out tensile creep experiments on unreinforced polypropylene after quenching the same from 80°C to 23°C as a function of aging time and stress. The long time creep data of Read and Tomlins (Read and Tomlins 1997) as a function of three aging times (7, 24 and 72 h) at tensile stress of 2.96 MPa is plotted in figure 7. Owing to aging during the course of creep $(t - t_w >> t_w)$, mere horizontal and vertical shifting is not sufficient to obtain the superposition. However transforming the data from the real time domain to the effective time domain leads to time – aging time superposition for $\mu$ =0.71 as shown in the inset of figure 7. In figure 8, we plot the long time creep data reported by Read and Tomlins (Read and Tomlins 1997) at five tensile stresses for an aging time of 24 h. It can be seen that compliance induced in the samples subjected to greater stresses is not just higher, but also evolves at a greater rate indicating superposition in the real time domian is not possible. We transfrom this data from the real time domain to the effective time domain, wherein the creep curve demonstrates self similar curvature. In the effective time domain, the compliance is plotted sgainst $[\xi(t) - \xi(t_w)] A / (\tau_0 \tau_m^{\mu-1})$. Since aging behavior is strongly affected by the stress field, $\mu$ depends on the applied stress while $A$, $\tau_0$ and $\tau_m$ are independent of stress. Consequently, a horizontal shifting of the creep curves is necessary to obtain the superposition in $\xi$ - domain. Upon horizontal shifting with factor $\alpha$, we get time – aging time – stress superposition as shown in figure 9. Similar to time – temperature superposition, in order to have abscissa of figure 9 represented by $[\xi(t) - \xi(t_w)] / \tau_0$, the horizontal shift factor $\alpha$ is given by: $\alpha = \tau_m^{\mu-\mu_R}$, where subscript $R$ represents value at the reference stress (2.96 MPa). In this comprehensive superposition we have also included time – aging time superposition showed in the inset of figure 7. In the inset of figure 9, we plot $\alpha$ as a function of $\mu - \mu_R$ on semi-logarithmic coordinates, the corresponding straight line indeed validates the relationship: $\alpha = \tau_m^{\mu-\mu_R}$, wherein slope of the same leads to estimation of $\tau_m$ =7.54×10$^{-4}$ s.

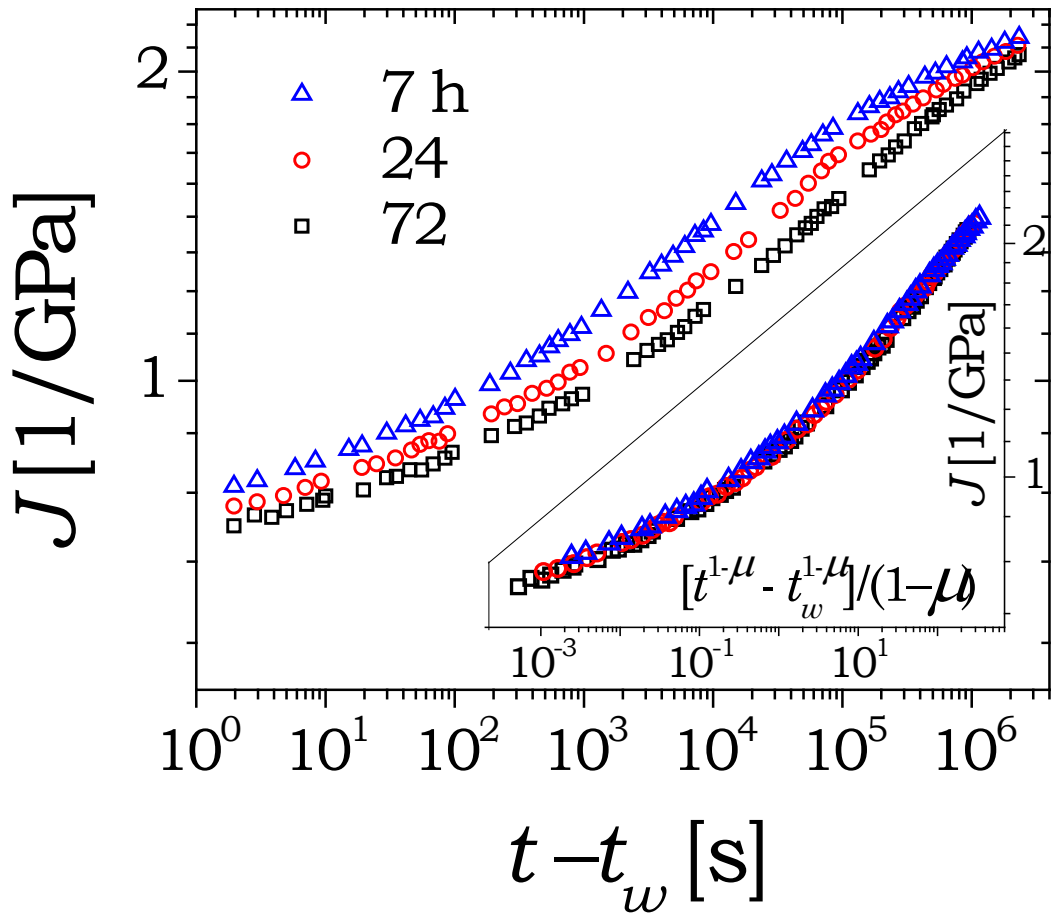

**Figure 7** The long time creep behavior of polypropylene under the tensile stress of 2.96 MPa applied at different aging times at 23°C. The aging time was measured after quenching the samples from 80°C to 23°C. The inset shows time – aging time superposition for the same data in the effective time domain for $\mu$=0.71. The data is adapted from Read and Tomlins (Read and Tomlins 1997)

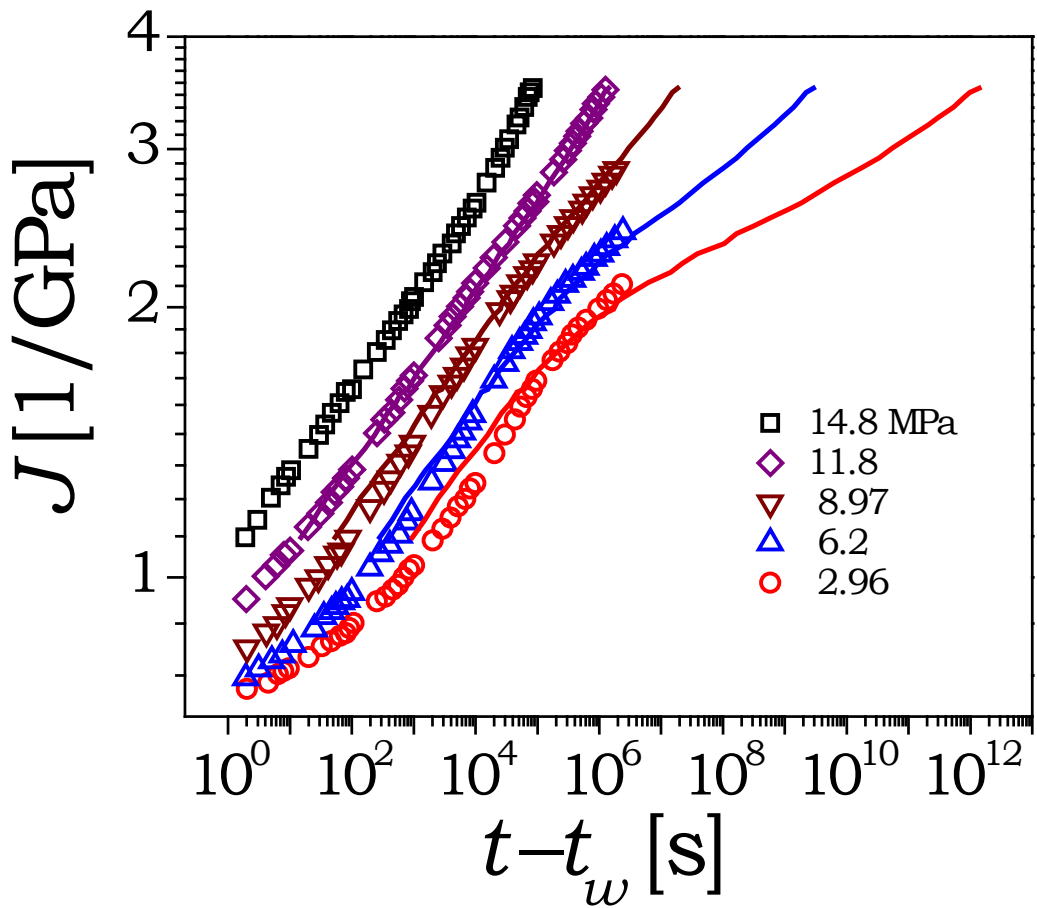

**Figure 8** The long time creep behavior of polypropylene under various tensile stresses applied at aging time of 24 h at 23°C. The symbols are the experimental data while the lines are the long time prediction of the creep data for the respective stresses obtained

from 14.8 MPa creep data. The data is adapted from Read and Tomlins (Read and Tomlins 1997).

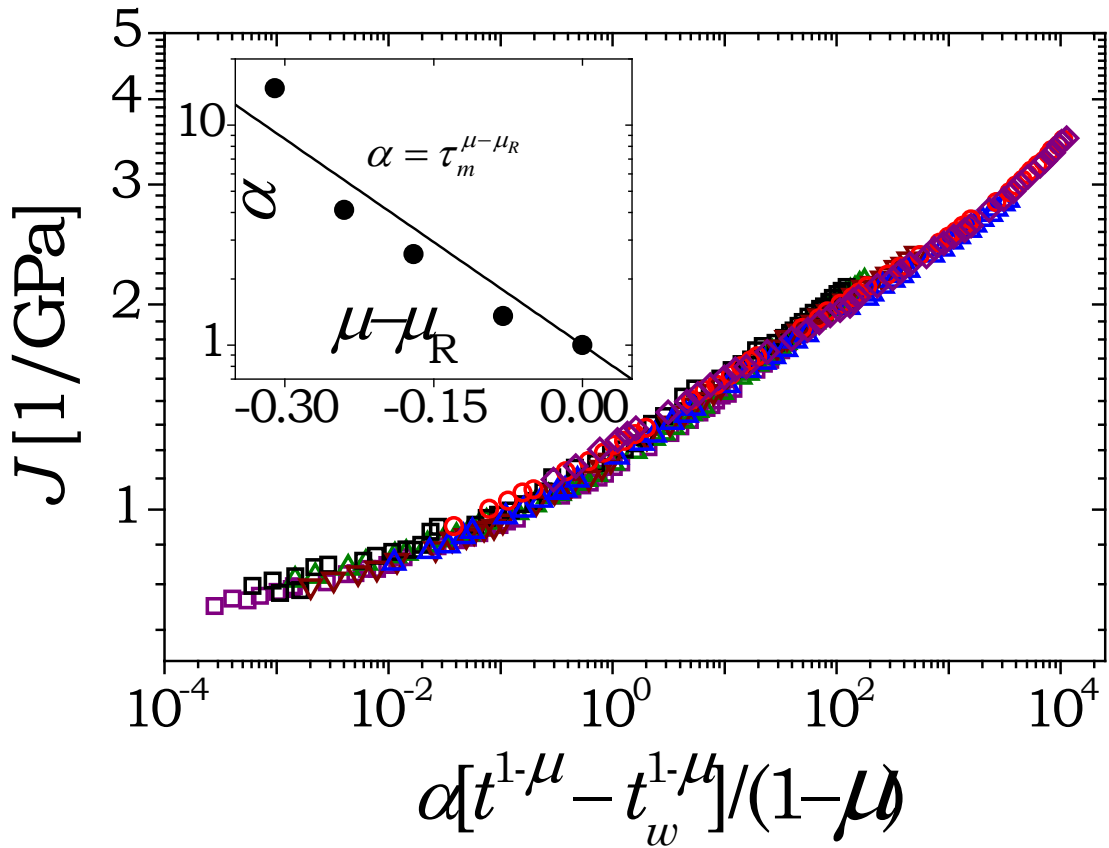

**Figure 9.** Time – aging time – stress superposition in the effective time domain for the data shown in figure 8 as well as that of in the figure 7. In order to obtain the superposition, values of $\mu$ for high to low stresses are respectively 0.4, 0.47, 0.54, 0.63 and 0.71. In the inset horizontal shift factor $\alpha$ is plotted as a function of $\mu - \mu_R$, where $\mu_R = 0.71$ is value of $\mu$ at the reference stress. The data is adapted from Read and Tomlins (Read and Tomlins 1997).

At higher stress not just the compliance is observed to higher (at any time) but also its rate of increase is also observed to greater. This suggests that the relaxation time and $\mu$ decreases with stress. Therefore, the creep data at higher stress can be employed to predict the creep behavior at lower stresses over longer duration. This can be easily done by using equation (9) and the corresponding prediction is shown in figure 8. It can be seen that the effective time domain approach leads to an excellent prediction of the creep compliance data. Furthermore, it can be seen that this procedure also leads to prediction of behavior at small stresses to very large time scales, demonstrating successful applicability of the effective time domain approch under application of different stresses.

In this work we have considered the effect of deformation field applied on the aging samples in terms of Boltzmann superposition principle expressed in the effective time domain. This offers a major advantage as the modified Boltzmann superposition principle allows analysis of more complicated flow fields. Read and Tomlins estimated compliance induced in the polypropylene at 23°C subjected to two step creep at two aging times. The tensile stress field applied to the sample is shown in figure 10a, while the consequent strain induced in the sample is shown in figure 10b. The expression of the effective time before applying the second step jump in the stress can be represented as:

$$\frac{[\xi(t) - \xi(t_w)]A}{\tau_0 \tau_m^{\mu_1 - 1}} = \left( \frac{t^{1-\mu_1} - t_w^{1-\mu_1}}{1 - \mu_1} \right) \qquad ...t_w \leq t \leq t_{w1}. \qquad (10)$$

Since the rate of aging depends strongly on the stress field, $\xi(t) - \xi(t_w)$ after applying the second stress step jump can be represented by:

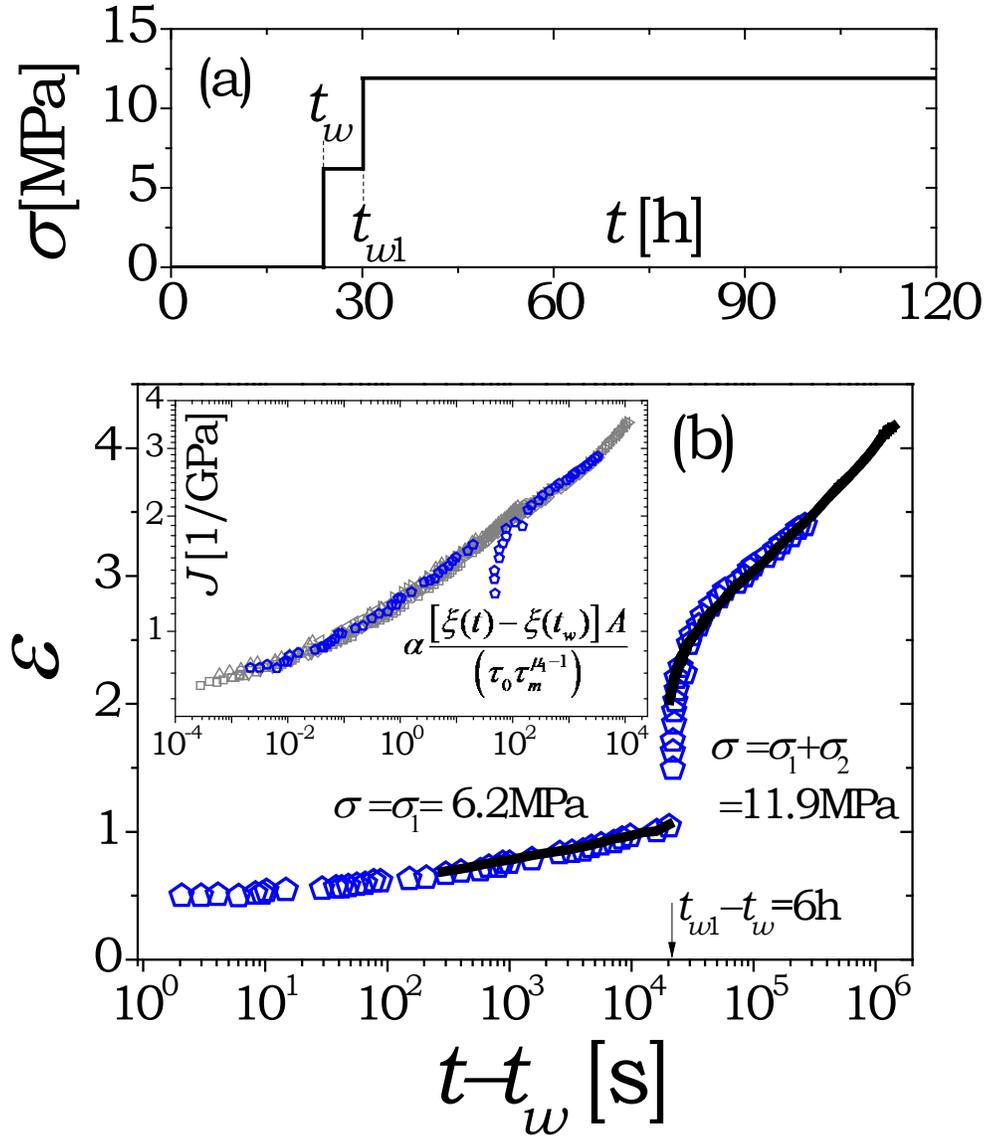

**Figure 10.** The plot (a) shows applied two-step tensile creep stress field wherein $\sigma_1$ =6.2 MPa was applied at $t_w$ =24 h, while additional stress of $\sigma_2$ =5.9 MPa was applied at $t_{w1}$ =30 h. The bottom figure shows tensile strain (symbols) induced in polypropylene after quenching the same to 23°C. The line is a prediction of strain obtained by transforming the creep data at 14.8 MPa to the real time domain. The inset in the bottom figure shows that two step evolution of compliance superposes very well on the master superposition (figure 9) in the effective time domain ($\mu_1$ =0.63 and $\mu_2$ =0.47). The data is adapted from Read and Tomlins (Read and Tomlins 1997).

$$\xi(t) - \xi(t_w) = \int_{t_w}^{t_{w1}} \frac{\tau_0 dt'}{A\tau_m^{1-\mu_1}t'^{\mu_1}} + \int_{t_{w1}}^{t} \frac{\tau_0 dt'}{A\tau_m^{1-\mu_2}t'^{\mu_2}} \qquad \ldots t > t_{w1}. \qquad (11)$$

In order to plot both the creep data, before and after applying the second step, in the effective time domain on the same chart, the abscissa for both must be the same. Consequently we divide equation (11) by $\tau_0 \tau_m^{\mu_1-1}/A$ to get the same left hand side as equation (10), which leads to:

$$\frac{[\xi(t) - \xi(t_w)]A}{\tau_0 \tau_m^{\mu_1-1}} = \left(\frac{t_{w1}^{1-\mu_1} - t_w^{1-\mu_1}}{1-\mu_1}\right) + \tau_m^{\mu_2-\mu_1}\left(\frac{t^{1-\mu_2} - t_{w1}^{1-\mu_2}}{1-\mu_2}\right) \qquad \ldots t > t_{w1}. \qquad (12)$$

In the inset of figure 10b we plot creep data shown in the figure 10b in terms of compliance with respect to $\alpha[\xi(t) - \xi(t_w)]A/(\tau_0 \tau_m^{\mu_1-1})$, where $\alpha$ is same as that of associated with 6.2 MPa in figure 9. Compliance is obtained by dividing the strain by respective creep stresses in both the steps. We also plot the time – aging time – stress superposition shown in figure 9 in grayscale. It can be seen that except the transient associated with transition in the step jump, both the creep data, before and after the step jump, fall right on top of the time – aging time – stress superposition. This result is remarkable because in order to plot the data in the effective time domain after the step jump, value of $\tau_m$ is necessary as shown in equation (12), which has been obtained independently by the fit to the shift factor data shown in the inset of figure 9 ($\tau_m$ =7.54×10$^{-4}$ s). This example therefore independently confirms the validity of the present approach. The very fact that the creep data beyond the step jump falls on the superposition suggests that inversion of the superposition from the effective time to the real time domain will lead to prediction of the long time creep behavior after the second step jump. We therefore transform the superposition or the creep data associated with the maximum stress from the effective time domain to the real time domain by inverting equation (12). The corresponding prediction is shown in figure 10b as a line. It can be seen that the present procedure gives a remarkable prediction of the strain induced in the material in the real time domain.

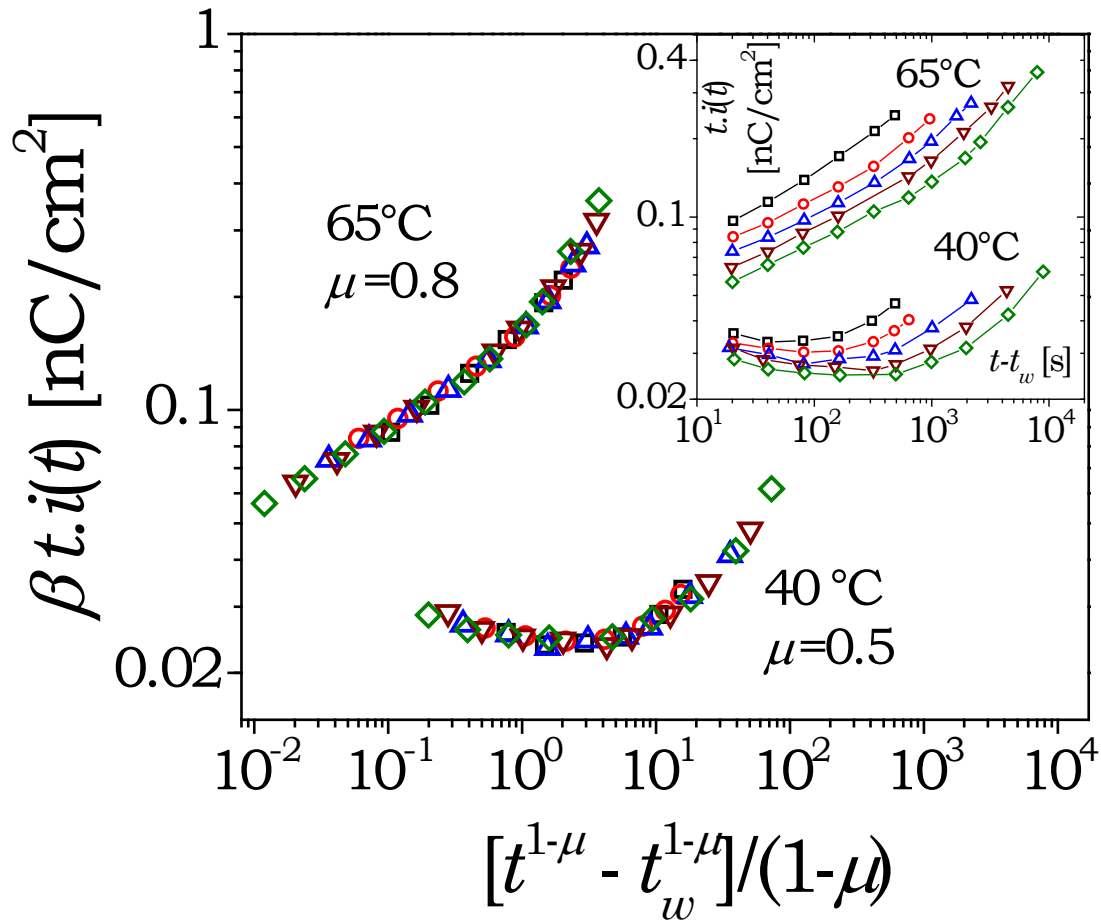

**Figure 11.** Time – aging time superposition resulting from Polarization experiment on rigid PVC quenched from 100°C to respective temperatures. The inset shows charge

accumulated in the sample upon application of step jump in DC voltage (3.5 kV/cm) at various aging times (squares: 720 s, circles: 1440 s, up triangles: 2700 s, down triangles: 5400 s, and diamonds: 10800 s). This data is adapted from Struik (Figure 51) (Struik 1978). In the main figure vertical shift factor $\beta$ is close to unity for all aging times at 65°C. At 40°C $\beta$ is 1, 0.92, 0.855, 0.8, 0.715 from higher to smaller aging times.

The applicability of the effective time domain approach discussed in this paper is not just limited to analyze the rheological behavior. This approach can be used to model, in principle, any time dependent processes. In figure 11 we plot results of a polarization experiment described by Struik (Figure 51) (Struik 1978). In these experiments rigid PVC quenched from 100°C to 65°C and 40°C have been subjected to DC voltage field strength of 3.5 kV/cm at various aging times. The charge accumulated in the polymer is plotted as a function of time elapsed since application of DC voltage in the inset of figure 11. It can be seen that lesser charge gets accumulated in the sample for DC voltage applied at higher aging times. Furthermore, the time over which the experiments have been carried out can be seen to be of the order of aging time. Consequently momentary superposition may not be possible by shifting the data in the real time as evident from the curvature of the data, particularly at 40°C. On the other hand, the experimental data plotted in the inset of figure 11 shows an excellent superposition upon transformation to the effective time domain, which we plot in figure 11. By using equation (9) this superposition can be easily converted to real time domain to predict charge accumulated in the samples over a long time period.

In addition to various examples mentioned above, it is important to note that the effective time domain approach offers great flexibility while analyzing the physical behavior of the materials in a perturbed state simultaneously undergoing aging. For example, a material following different rates of enhancement of relaxation time or relaxation time dependence having different functional forms subsequent to each other in the same experiment can be easily incorporated into effective time domain approach using equation (3). Furthermore, since effective time domain approach is used in terms of linear viscoelastic framework, complicated but known deformation histories within the purview of linear viscoelasticity can be taken into account while analyzing stress/strain induced in the material. We believe that both these points offer significant benefit while analyzing the aging behavior of the amorphous polymeric compounds. We hope that various features of the effective time domain framework presented in this work will initiate further experimental and theoretical investigations on variety of polymeric systems under different deformation histories.

## V. Conclusions

The amorphous polymeric materials are known to undergo physical aging wherein free volume relaxation causes temporal enhancement of the relaxation time. As a result, owing to time dependency associated with various properties of the aging materials, prediction of the long time physical behavior poses challenges. In this work we apply the effective time domain approach to predict the long time physical behavior of aging amorphous polymeric materials. In the effective time domain the real time is normalized by the time dependent relaxation time such that the relaxation dynamics remains invariant of effective time. In this approach rheological (or physical) data associated with aging material in the real time domain is transformed to the effective time domain. Since time dependency ceases in the effective time domain, time – aging time superposition, time – temperature superposition and time – stress superposition facilitate prediction of the long time physical behavior of amorphous polymeric materials. Interestingly, unlike the conventional approach by Struik to predict the long time behavior, the effective time domain approach can account for aging that occurs over the duration of experiment. Consequently the information available over the longer duration of experiments can be utilized in the effective time domain approach unlike that of in the conventional procedure. We feel that this is a major advantage that effective time domain approach offers compared to the conventional procedure.

We apply effective time domain approach to the experimental data on amorphous polymers due to Struik (Struik 1978). We observe that data collapses to form a master curve in the effective time domain even in the limit of process time (time over which impetus is applied) >> aging time (time at which impetus is applied). In addition, the experimental data on aging amorphous polymers successfully demonstrates application of time – temperature superposition in the effective time domain. Consequently 9 experiments, each performed over a time scale of around 30 hours at three aging times and three temperatures each, lead to prediction of creep behavior over extraordinary time scales (around 6 orders of magnitude greater than the aging time). We believe that such predictive capacity can only be obtained through the effective time domain approach. In another example, we apply the effective time domain approach to the momentary data (in the limit of process time << aging time) and demonstrate that the present approach leads to better prediction of the long time behavior than the conventional approach. This suggests even in the momentary limits aging during the short experimental time scales cannot be ignored.

We further analyze creep data obtained at different stresses and demonstrate existence of time – stress superposition in the effective time domain. Interestingly the creep data associated with two-step stress change also obeys the superposition. Consequently transformation of the superposition from the effective time domain to the real time domain leads to excellent prediction of the single step as well as the two step creep behavior of aging polymeric material. Finally we demonstrate successful application of effective time domain approach to experimental data on charge accumulated in the amorphous polymer when subjected to DC voltage at different aging times. Various examples studied in this paper and the associated discussion clearly establish that effective time domain approach can serve as an important tool to analyze aging amorphous polymeric materials and offers many advantages over the presently available procedures.

**Acknowledgement:** Financial support from Department of Atomic Energy – Science Research Council, Government of India is greatly acknowledged.

**References:**

Arnold JC, White VE (1995) Predictive models for the creep behaviour of PMMA. Materials Science and Engineering: A 197: 251-260. DOI 10.1016/0921-5093(95)09733-3

Awasthi V, Joshi YM (2009) Effect of temperature on aging and time–temperature superposition in nonergodic laponite suspensions. Soft Matter 5: 4991–4996

Baldewa B, Joshi YM (2012) Delayed Yielding in Creep, Time - Stress Superposition and Effective Time Theory for a soft Glass. Soft Matter 8: 789-796

Bird RB, Armstrong RC, Hassager O (1987) Dynamics of Polymeric Liquids, Fluid Mechanics. Wiley-Interscience, New York

Bradshaw RD, Brinson LC (1997) Physical aging in polymers and polymer composites: An analysis and method for time-aging time superposition. Polymer Engineering & Science 37: 31-44. DOI 10.1002/pen.11643

Callen HB (1985) Thermodynamics and an introduction to thermostatistics. John Wiley & Sons, New York

Cloitre M, Borrega R, Leibler L (2000) Rheological aging and rejuvenation in microgel pastes. Phys Rev Lett 85: 4819-4822

Denman HH (1968) Time-Translation Invariance for Certain Dissipative Classical Systems. American Journal of Physics 36: 516-519. DOI 10.1119/1.1974957

Derec C, Ducouret G, Ajdari A, Lequeux F (2003) Aging and nonlinear rheology in suspensions of polyethylene oxide-protected silica particles. Phys Rev E 67: 061403

Donth E (2001) The Glass Transition. Springer, Berlin

Dorigato A, Pegoretti A, Kolařík J (2010) Nonlinear tensile creep of linear low density polyethylene/fumed silica nanocomposites: Time-strain superposition and creep prediction. Polymer Composites 31: 1947-1955. DOI 10.1002/pc.20993

Fielding SM, Sollich P, Cates ME (2000) Aging and rheology in soft materials. J Rheol 44: 323-369

Guo Y, Bradshaw RD (2009) Long-term creep of polyphenylene sulfide (PPS) subjected to complex thermal histories: The effects of nonisothermal physical aging. Polymer 50: 4048-4055. DOI 10.1016/j.polymer.2009.06.046

Guo Y, Wang N, Bradshaw RD, Brinson LC (2009) Modeling mechanical aging shift factors in glassy polymers during nonisothermal physical aging. I. Experiments and KAHR-ate model prediction. Journal of Polymer Science Part B: Polymer Physics 47: 340-352. DOI 10.1002/polb.21643

Gupta R, Baldewa B, Joshi YM (2012) Time Temperature Superposition in Soft Glassy Materials. Soft Matter 8: 4171

Hodge IM (1994) Enthalpy relaxation and recovery in amorphous materials. J Non-Cryst Solids 169: 211-266. DOI 10.1016/0022-3093(94)90321-2

Hodge IM (1995) Physical aging in polymer glasses. Science 267: 1945-1947

Hopkins IL (1958) Stress relaxation or creep of linear viscoelastic substances under varying temperature. Journal of Polymer Science 28: 631-633. DOI 10.1002/pol.1958.1202811817

Jazouli S, Luo W, Bremand F, Vu-Khanh T (2005) Application of time-stress equivalence to nonlinear creep of polycarbonate. Polymer Testing 24: 463-467

Joshi YM, Reddy GRK (2008) Aging in a colloidal glass in creep flow: Time-stress superposition. Phys Rev E 77: 021501-021504

Kolařík J (2003) Tensile creep of thermoplastics: Time-strain superposition of non-iso free-volume data. Journal of Polymer Science, Part B: Polymer Physics 41: 736-748. DOI 10.1002/polb.10422

Kolarik J, Pegoretti A (2006) Non-linear tensile creep of polypropylene: Time-strain superposition and creep prediction. Polymer 47: 346-356

Kovacs AJ, Aklonis JJ, Hutchinson JM, Ramos AR (1979) Isobaric volume and enthalpy recovery of glasses 2. Transparent multi-parameter theory. J Polym Sci Pt B-Polym Phys 17: 1097-1162. DOI 10.1002/pol.1979.180170701

Larson RG (1999) The Structure and Rheology of Complex Fluids. Clarendon Press, Oxford

McKenna GB (1994) On the physics required for prediction of long term performance of polymers and their composites. J Res Natl Inst Stand Technol 99: 169

McKenna GB (2003) Mechanical rejuvenation in polymer glasses: fact or fallacy? J Phys: Condens Matter 15: S737–S763

McKenna GB, Narita T, Lequeux F (2009) Soft colloidal matter: A phenomenological comparison of the aging and mechanical responses with those of molecular glasses. Journal of Rheology 53: 489-516

Narayanswamy OS (1971) Model of structural relaxation in glass. Journal of the American Ceramic Society 54: 491-498. DOI 10.1111/j.1151-2916.1971.tb12186.x

Negi AS, Osuji CO (2009) Dynamics of internal stresses and scaling of strain recovery in an aging colloidal gel. Phys Rev E 80: 010404

Nicholson LM, Whitley KS, Gates TS (2000) Molecular Weight Effects on the Viscoelastic Response of a Polyimide. NASA Langley Technical Report Server

O'Connell PA, McKenna GB (1997) Large deformation response of polycarbonate: Time-temperature, time-aging time, and time-strain superposition. Polym Eng Sci 37: 1485-1495

O'Connell PA, McKenna GB (2002) The non-linear viscoelastic response of polycarbonate in torsion: An investigation of time-temperature and time-strain superposition. Mechanics Time-Dependent Materials 6: 207-229

Read BE, Dean GD, Tomlins PE, Lesniarek-Hamid JL (1992) Physical aging and creep in PVC. Polymer 33: 2689-2698. DOI 10.1016/0032-3861(92)90439-4

Read BE, Tomlins PE (1997) Time-dependent deformation of polypropylene in response to different stress histories. Polymer 38: 4617-4628. DOI 10.1016/S0032-3861(96)01085-3

Read BE, Tomlins PE, Dean GD (1990) Physical ageing and short-term creep in amorphous and semicrystalline polymers. Polymer 31: 1204-1215. DOI 10.1016/0032-3861(90)90209-H

Reddy GRK, Joshi YM (2008) Aging under stress and mechanical fragility of soft solids of laponite. Journal of Applied Physics 104: 094901

Rodriguez GF, Kenning GG, Orbach R (2003) Full aging in spin glasses. Phys Rev Lett 91: 037203. DOI 10.1103/PhysRevLett.91.037203

Schoeberle B, Wendlandt M, Hierold C (2008) Long-term creep behavior of SU-8 membranes: Application of the time-stress superposition principle to determine the master creep compliance curve. Sensors and Actuators, A: Physical 142: 242-249. DOI 10.1016/j.sna.2007.03.017

Shahin A, Joshi YM (2011) Prediction of long and short time rheological behavior in soft glassy materials. Phys Rev Lett 106: 038302

Shahin A, Joshi YM (2012) Hyper-Aging Dynamics of Nanoclay Suspension. Langmuir 28: 5826-5833. DOI 10.1021/la205153b

Shaukat A, Sharma A, Joshi YM (2010) Time-aging time-stress superposition in soft glass under tensile deformation field. Rheologica Acta 49: 1093-1101. DOI 10.1007/s00397-010-0469-2

Shi X, Mandanici A, McKenna GB (2005) Shear stress relaxation and physical aging study on simple glass-forming materials. Journal of Chemical Physics 123: 174507. DOI 10.1063/1.2085050

Sibani P, Kenning GG (2010) Origin of end-of-aging and subaging scaling behavior in glassy dynamics. Physical Review E 81: 011108

Struik LCE (1978) Physical Aging in Amorphous Polymers and Other Materials. Elsevier, Houston

Tomlins PE, Read BE (1998) Creep and physical ageing of polypropylene: a comparison of models. Polymer 39: 355-367. DOI 10.1016/S0032-3861(97)00258-9

Tomlins PE, Read BE, Dean GD (1994) The effect of temperature on creep and physical ageing of poly(vinyl chloride). Polymer 35: 4376-4381. DOI 10.1016/0032-3861(94)90095-7

Zheng SF, Weng GJ (2002) A new constitutive equation for the long-term creep of polymers based on physical aging. European Journal of Mechanics - A/Solids 21: 411-421. DOI 10.1016/s0997-7538(02)01215-9